\documentclass{aa} 

\usepackage{txfonts}
\usepackage{graphicx}

\usepackage{natbib}
\bibpunct{(}{)}{;}{a}{}{,}
\title{On the colour-colour properties of the Extremely Red Objects}
\titlerunning{On the colour-colour properties of the EROs}

\author{Stefan Bergstr{\"o}m\inst{1} \and Tommy Wiklind\inst{2}}
\authorrunning{S.\ Bergstr{\"o}m \and T.\ Wiklind}

\institute{Department of Astronomy and Astrophysics, Centre for Astrophysics and 
  Space Science, Chalmers University of Technology, SE-439 92 Onsala, Sweden, 
  \email{stefan@oso.chalmers.se} \and 
  European Space Agency, Space Telescope Science Institute, 3700 San Martin Drive, 
  Baltimore, MD 21218, USA, \email{wiklind@stsci.edu}}

\offprints{Stefan Bergstr\"om,\\ \email{stefan@oso.chalmers.se}}

\date{Received $<$date$>$ / Accepted $<$date$>$}

\abstract{The colours of the galaxy class known as Extremely Red Objects
  (EROs; $R-K>5$) are considered to be consistent with two distinct galaxy
  populations at high redshift: evolved ellipticals or young dusty
  starbursts. In this paper the properties of EROs, spanned by the
  five photometric bands $RIJHK$, are investigated as to the
  possibility to distinguish between these two galaxy populations
  using only broad band photometry. The broad band colours of
  elliptical and starburst galaxies at redshifts up to $5$ are computed from
  synthetic spectra obtained using the spectral evolution synthesis programme
  P\'EGASE.2. Two initial mass functions and a range of metallicities and
  extinctions are used. In order to be extremely red in the redshift range
  considered, the evolution of the $R-K$ colour sets the requirement that
  ellipticals have to be less than $7-8$ Gyr old, and that the starbursts
  must have colour excesses of $E(B-V)>1$, as derived from the nebular
  emission lines. In investigating the overlap in the  different colour-colour
  planes as a function of redshift, it is found that the planes formed from
  permutations of the same three filters exhibit very similar overlap
  characteristics.  In colour-colour planes formed within such triplets one of the
  filters will serve as a ``pivot'' band against which the two other bands are
  compared. The configuration where this pivot band lies between the other two
  bands has the best performance as a discriminator among the three possible
  configurations. A consistent behaviour cannot be found among the
  configurations formed by permuting four filters.  The minimal filter
  configuration $R-H$ vs.\ $H-K$ is found to be the very best discriminator,
  working as such up to redshift $2.9$.  
}

\begin{document} 

\maketitle

\section{Introduction} 

The optically faint high-redshift galaxy class known as Extremely Red
Objects, or EROs, has since its discovery \citep{err1988} been
extensively studied. The EROs are considered to be one of the more enigmatic
galaxy classes since their faintness makes it very difficult to
classify them, either spectroscopically or morphologically. Although their extremely
red colours, $R-K > 5$, narrow down the possibilities 
to essentially ellipticals and dusty starburst galaxies at high
redshifts, the colour degeneracy between these two very different
galaxy populations prevents the determination of their morphologic
types. Understanding the nature of the EROs is of great importance
since they can provide a direct observational test of theories of
galaxy formation \citep[e.g.][]{cimatti2001}: the hierarchical
clustering picture
\citep[e.g.][]{white_frenk_1991,kauffmann_white_guiderdoni_1993}
versus the monolithical collapse picture
\citep[e.g.][]{eggen_lynden-bell_sandage_1962,larson_1975}.  
\par
Due to their faintness, only a handful EROs have reliable redshift
estimates in combination with a determination of the morphologic
type. These individual cases divide evenly between starbursts
\citep[e.g.][]{cimatti_et_al_1998,dey_et_al_1999,afonso_et_al_2001,smith_et_al_2001,frayer_et_al_2003}
and ellipticals
\citep[e.g.][]{spinrad_et_al_1997,soifer_et_al_1999,stiavelli_et_al_1999,saracco_et_al_2003},
and do not allow any conclusion to be drawn as to what galaxy
type the majority of the EROs belongs to. The individually
studied elliptical EROs are located in a narrow redshift range, $1.2 <
z < 1.8$, whereas the individually classified starburst systems span a
much larger redshift range. 
\par
Surveys have shown that the EROs comprise a heterogeneous galaxy
population \citep{thompson1999,cimatti1999,moriondo2000}, although the elliptical case has been the
stronger one supported by in particular the detected clustering of the EROs
\citep{daddi_et_al_2000,moustakas_somerville_2002}. However, the sample of EROs compiled by
\citet{daddi_et_al_2002} contains similar fractions of star-forming
and evolved galaxies, with only the evolved galaxies showing strong
clustering. In later surveys, the heterogeneity of the ERO population
has been confirmed
\citep[e.g.][]{mannucci_et_al_2002,smail_et_al_2002,smith_et_al_2002a,yan_thompson_2003,
moustakas_et_al_2003}
and the picture is further complicated by the finding by
\citet{smith_et_al_2002b} of an early-type disc ERO at $z=1.6$. 
\par
Taking the properties of the ERO sample (complete to $K_s \leq 19.2$) obtained by
\citet{cimatti_et_al_2002} as ``typical'', the ERO population consists
of similar fractions of ``old'' and ``star-forming'' galaxies. The
average ``old'' ERO is concluded to be at least $3$ Gyr old, being
formed beyond $z = 2.4$, if the metallicity is solar. The average
``star-forming'' ERO agrees with a starburst spectrum having a
\citet{calzetti2000} $E(B-V)_\mathrm{star}$ of $\sim\!0.8$, i.e.\ an
extinction for the nebular emission lines of $E(B-V) = 1.8$. If the
contribution from an assumed underlying old stellar population is
accounted for, the average reddening can be decreased to
$E(B-V)_\mathrm{star}\sim 0.7$. Using other synthetic spectra, the
extinction is found to be in the range $0.6 < E(B-V)_\mathrm{star} < 1.1$.
\par
\par
Since EROs are by their definition very weak at optical wavelengths, spectroscopy
is not a viable method for classifying large numbers of EROs. Instead, attempts
have been made to use broad band colours. \citet[][ hereafter PM00]{pm2000}
found that EROs in the redshift ranges $1 \leq z \leq 2$ and $2 \leq z \leq 2.5$
fall in different regions in both $I-K$ vs.\ $J-K$ and $R-K$ vs.\ $J-K$
plots for sources in the lower redshift range, and in $H-K$ vs.\ $J-K$ plots for
sources in the higher redshift range. This result is robust and valid for a variety
of metallicities, dust contents, star formation histories and
redshifts of formation. 
\par
\citet{pierini_et_al_2003} recently extended the study of the colour-colour
method for classifying EROs. The study is similar to that of PM00
but includes dusty post-starburst galaxies as well as the nominal
dust-free evolved stellar populations (old ellipticals) and dusty
starbursts. \citeauthor{pierini_et_al_2003} also studied the impact of various 
dust distributions within the model galaxies. While the dust distribution
can change the attenuation properties, it did not have a large impact
on the overall results, which were similar to those of PM00.
The presence of dusty post-starburst systems tends to make the
separation of old, dust-free systems and dusty starburst systems
less distinct.
\par
In this work, the possibility to use broad band colours to
distinguish between the elliptical and dusty starburst galaxy
populations among the EROs at redshifts below five is further pursued, and all possible
colour-colour configurations of the five broad band filters 
$RIJHK$ are investigated. The study is based on models of the
elliptical and dusty starburst populations, where synthetic galaxy
spectra are obtained using the spectral evolution synthesis programme
P\'EGASE.2 \citep{frv1997, pegase}. In particular, colour-colour
combinations are searched for where the two galaxy populations do not
overlap over a large redshift range. The intention is to
study to what extent photometric classifications of EROs can be used.
The adopted cosmology is $H_0 = 70$ km$\,$s$^{-1}\,$Mpc$^{-1}$,
$\Omega_\mathrm{m} = 0.3$ and $\Omega_\Lambda = 0.7$, thus yielding an
age of the Universe of $13.5$ Gyr.
\par
The paper is organised as follows. The properties of the different
galaxy models are presented in Sect.~\ref{galaxy_models_section}. In
Sect.~\ref{method_section} the methods used are accounted for, which
include determining which models produce EROs considering redshift and
galaxy age, and investigating the overlap between the colours of the
galaxy populations in different colour-colour diagrams. The results are
presented and discussed in Sect.~\ref{results_section}, and 
Sect.~\ref{summary_section} concludes.

\section{The galaxy models}
\label{galaxy_models_section}

In general there are two distinct galaxy populations that the EROs are
considered to belong to: ellipticals and dusty starbursts. The elliptical
galaxies consist of an old stellar population and have none or small amounts
of gas and dust. Hence the colours of these galaxies are determined by the old 
stellar population which produce a large 4000 \AA\ break in the spectral energy
distribution (SED). The colours of ellipticals can therefore be very red,
especially if the galaxies are located at high redshift. The other galaxy 
population producing EROs is the dusty starbursts. These galaxies contain much 
gas and dust and have high star formation rates (SFRs). The light of the early 
stellar population is scattered and absorbed by the dust, which heats up and 
subsequently re-radiates the energy in the far-infrared. Since this extinction 
is greater at shorter wavelengths, the initially blue SED is reddened. 
Furthermore, a redshifted SED will be redder than the restframe counterpart. 
\par
To model these galaxy populations, the spectral synthesis evolution programme
P\'EGASE.2\footnote{P\'EGASE.2 is made publicly available at the web site\\
  \mbox{\texttt{http:$/\!/$www.iap.fr$/$pegase$/$}}.} \citep{frv1997, pegase} is
used. The programme computes models of 
galactic spectra in the range $91$ \AA\ to $160~\mu$m assuming
different star formation scenarios, metallicities etc. P\'EGASE.2 uses
evolutionary tracks of mainly the Padova group (see \citealp{pegase}
for details), and synthetic spectra are used when observational libraries are not available. For the
nebular emission, only case B recombination is taken into account when
computing the hydrogen line fluxes. For other lines, typical observed H$\beta$
ratios are used. The programme allows for a consistent metallicity
evolution using the \citet{ww1995} models of SNII ejecta.
\par
The elliptical galaxies are modelled as galaxies that after an
instantaneous burst of star formation evolve passively, also
known as a simple stellar population (SSP) model. (This is the
(instantaneous-)burst population of
\citealp{bruzual_charlot_1991,bruzual_charlot_1993}.) After
the burst of star formation the galaxies are assumed to contain no
interstellar gas or dust, which is equivalent to assuming the on-set
of a strong galactic wind. The galaxies are considered as ellipticals
100 Myr after the starburst, and are then let to evolve for
a Hubble time. The time grid consists of $128$ epochs with a spacing
of about $5\%$ of the galaxy age. 
\par
The starburst galaxies are modelled as galaxies with a constant star
formation during $100$ Myr. The time step is $1$ Myr. The extinction
in the starbursts is modelled as a uniform screen of dust and gas and
follows the \citet[][ hereafter C00]{calzetti2000} obscuration law
derived for starbursts. According to the work by C00, the stellar
continuum and the nebular emission lines suffer from different
extinctions, due to the different stellar environments. The extinction
of the stellar continuum is proportional to the extinction of the
nebular emission lines as given by the empirically derived relation  
\begin{equation}
  E(B-V)_\mathrm{stars} = 0.44\,E(B-V)_\mathrm{gas}.
\end{equation}
Throughout this paper the extinction will be parameterised using the
colour excess of the nebular emission lines, and the subscript
\emph{gas} is dropped (e.g.\ C00).
\par
The C00 extinction law expresses the selective absorption, not
total. The shape of the extinction law depends on the properties of the dust grains
causing the extinction, whereas the \emph{amount} of extinction can
vary significantly. Here, the extinction is treated as a screen
through which the radiation from the galaxies passes, and the amount
of extinction can be varied by increasing the thickness of the
screen. The colour excess required to produce an ERO of a particular
galaxy at a given redshift is therefore only indicative of the amount
of extinction needed. The total column density of dust, for example,
will be dependent on the geometry of the stars and the obscuring
medium. 
\par
The choice of SFR in the starbursts requires some consideration as it
influences the SED in two ways. (\textbf{i}) The SFR provides a scaling
factor of the flux of the spectrum. This scaling property does not
have any impact on the colours as these are derived from the shape of
the spectrum, not its absolute value. This is true also in the
presence of nebular emission lines, as the continuum flux and emission
line flux (through the flux of ionising photons) both are proportional
to the number of stars produced. This has the result that two galaxies
of the same metallicity, IMF and temporal \emph{shape} of the star
formation history will have the same colours regardless of the
magnitude of the SFR. (\textbf{ii}) The SFR will influence the rate at
which the galaxies are enriched: the higher the SFR, the faster the
enrichment. This will cause a major change in the colours. Hence, of
these two effects it is only the latter, where the metallicity is
increased, that has a significant effect on the colours. 
\par
Since a consistent evolution of the metallicity is used in the
modelling of the galaxy spectra, the metallicities will change in a galaxy
experiencing an extended period of star formation (i.e. the starburst model).
Following the line of reasoning sketched out in the previous
paragraph, the SFRs cannot be too high if the range of 
metallicities is to be kept in the modelling.
Hence, the value of the SFR is chosen to achieve a well
controlled range of metallicities. The metallicity grid consists of
five metallicities, $0.02$, $0.2$, $0.4$, $1$ and $2.5$ Z$_\odot$,
which in the passive evolution of the ellipticals do not change, but
in the case of the starbursts, these are initial values. 
\par
The SFR chosen to ensure such a slow enrichment is $5\times 10^{-11}
M_\mathrm{galaxy}$ yr$^{-1}$. A galaxy with such an SFR would
not be regarded as a starburst galaxy since only $0.5\%$ of the gas
mass have been used to form stars after $100$ Myr. However, since the
magnitude of the SFR does not influence the colours of galaxies of the
same metallicity and stellar population, as explained above, this
low-SFR model is equivalent to models having a more ``realistic'' SFR.  
Considering the metallicity in the ISM, an initial solar metallicity
has been increased to $1.0050$ Z$_\odot$, i.e. by $0.5$\%, after $100$
Myr for a starburst of Salpeter IMF. The enrichment is stronger the
lower the initial metallicity, and the highest increase of metallicity
occurs for the starbursts of initial metallicity $0.02$ Z$_\odot$\ and
Salpeter IMF; after $100$ Myr, the metallicity have increased by $\sim\!25\%$.  
\par
The initial mass functions (IMFs) used in the modelling are the \citet{salpeter1955} and  
the ``present-day'' \citet[][ equation (6)]{kroupa_2001} IMFs in the
mass range $0.09\leq$ M$/$M$_\odot \leq 120$. The Kroupa IMF is mainly based on the
power-law compilation by \citet{scalo1998} but the effects of unresolved
binaries are also taken into account. This IMF differs from the Salpeter IMF
below $1$~M$_\odot$ and is included to increase the generality of the
results. The adopted mass limits are consistent with the stellar tracks used
by P\'EGASE.2. The parameters common to all the models are shown in \mbox{Table
  \ref{otherprops}}.
\par
\begin{figure}
\centering
\resizebox{\hsize}{!}{\includegraphics{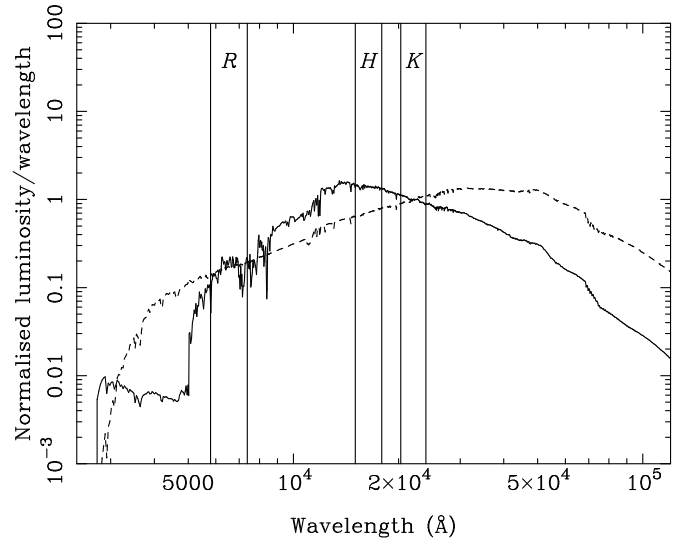}}
\caption{Spectral energy distributions of galaxies
  at $z = 2$ having Salpeter IMF and solar 
  metallicity, normalised in the $K$ band. The \emph{solid line} is
  the SED of a $1$ Gyr old elliptical galaxy; the \emph{dashed line} is
  the SED of a $100$ Myr old starburst galaxy having $E(B-V) = 2$. Shown
  are also the wavelength regions covered by the $R$, $H$ and $K$ filters.}
\label{compare}
\end{figure}
\begin{figure*}
\centering
\resizebox{17cm}{!}{
\includegraphics[width = 88mm]{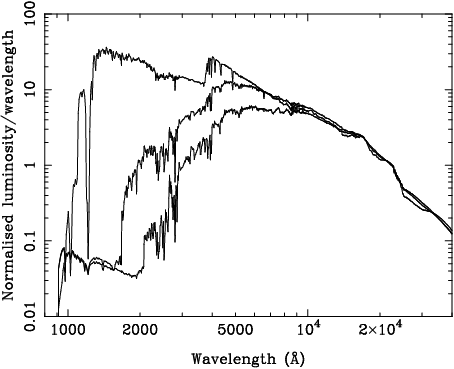}
\hspace{3em}
\includegraphics[width = 88mm]{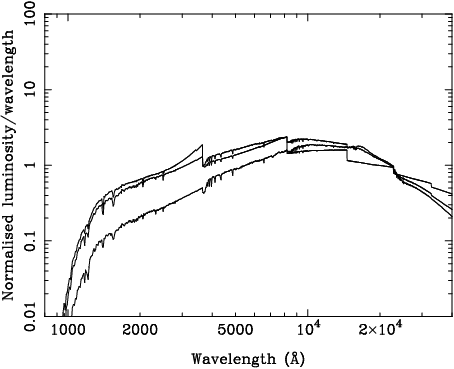}}
\caption{Spectral energy distributions of galaxies having Salpeter IMF and solar
  metallicity, normalised in the $K$ band. \emph{Left panel:} a $200$ Myr, $1$ Gyr and $5$ Gyr old 
  elliptical. \emph{Right panel:} a new-born, $10$
  and $100$ Myr old starburst galaxy having $E(B-V) = 2$. Both galaxy
  types become redder with age.}
\label{spectra}
\end{figure*}

Model spectra of a $1$ Gyr old elliptical and a $100$ Myr old starburst
galaxy of $E(B-V) = 2$ are shown in Fig.~\ref{compare}. The galaxies have
Salpeter IMF and solar metallicity. Inlaid are the wavelength regions
of the $R$, $H$ and $K$ bands. In Fig.~\ref{spectra}, spectra of galaxies
of ellipticals $200$, $1000$ and $5000$ Myr old (left panel) and of
starbursts $0$, $10$ and $100$ Myr old having $E(B-V) = 2$ (right panel) are
shown.

\begin{table}
\centering
\caption{P\'EGASE.2 parameters common to all the models}
\label{otherprops}
\begin{tabular}{ll}
\hline
\hline
Property                         & Value\\
\hline
Supernova ejecta                 & \citet{ww1995} \\
                                 & \qquad model B \\
Infall                           & No \\
Stellar winds                    & Yes \\
Fraction of close                &     \\
\qquad binary systems            & 0.05\\
Metallicity evolution            & Consistent\\
Formation of substellar          &   \\
\qquad objects                   & No\\
Nebular emission                 & Yes\\
Extinction                       & No\\
\hline
\end{tabular}
\end{table}

\section{The method}
\label{method_section}

Two questions are addressed in the present work: (\textbf{i})
When can ellipticals and starbursts be EROs? (\textbf{ii}) Is
it possible to distinguish between the extremely red ellipticals and
dusty starbursts merely from their broad band colours? A first attempt
to answer the second question was made by \citet{ss2000}. In a similar
study, PM00 were able to separate old elliptical galaxies
from dusty starbursts using $I-K$ vs.\ $J-K$ and $R-K$ vs.\ $J-K$ plots in the
redshift range $1 \leq z \leq 2$, and also $H-K$ vs.\ $J-K$ in the
range $2 \leq z \leq 2.5$.
\par
In the work presented here, the ability to distinguish
between the considered galaxy types are investigated for all the $45$
possible combinations of the $RIJHK$ filters and for all plausible redshift
ranges. This section describes the methods used to investigate the colour
evolution and the colour plane properties. 
\par
The fluxes at different redshifts are computed from the spectra obtained from
P\'EGASE.2 using the standard relation
\begin{equation}
  F(\lambda_0) = \displaystyle\frac{L(\lambda_1) (1+z)}{4\pi D_L^2},
\end{equation}
where $\lambda_0 = (1+z)\lambda_1$ and $D_L$ is the luminosity distance.
The magnitudes are then computed in the Vega system using the well-documented
filter functions of \citet{bb1988} and \citet{bessell1990}
founded on the Cousins system. The filter response functions are
extended to $1500$ data points using cubic splines. This filter system
is very similar to the filters used in the EIS-Deep Survey, the SUSI
$RI$ and SOFI $JHK$ filters at the NTT\footnote{Filter functions
  achieved from the Filter Catalogue at the ESO La Silla home page,
  \mbox{\texttt{http:$/\!/$www.ls.eso.org}}.}, so any conclusions made
in the following will not be altered significantly if the latter
system is used.
 
\begin{table}
\centering
\caption{The properties of the normalised (unity peak transmission) Bessell
  filters. The width is defined as the area of the filter response
  function.}
\label{filterprop}
\begin{tabular}{llll}
\hline
\hline
Band & $\lambda_\mathrm{mean}$ (\AA) & Width (\AA) & $F_0 $ (W$\,$m$^{-2}\,$\AA$^{-1}$) \\
\hline
$R$ & $\phantom{0}6586$ &    $1581$ & $2.216\times 10^{-12} $ \\
$I$ & $\phantom{0}8060$ &    $1495$ & $1.154\times 10^{-12} $ \\
$J$ &    $12368$ &    $2023$ & $3.256\times 10^{-13} $ \\
$H$ &    $16466$ &    $2855$ & $1.181\times 10^{-13} $ \\
$K$ &    $22119$ &    $3660$ & $4.110\times 10^{-14} $ \\
\hline
\end{tabular}
\end{table}

\subsection{Colour evolution}
\label{erosidentsec}

Two approaches are used to find galaxies that might produce EROs under certain
circumstances. The first approach is to find the lowest redshift required to
make a galaxy of a certain age extremely red. The redshift where $R-K=5$ will
hereafter be called $z_\mathrm{ERO}$.
This procedure imposes a cosmology consistency criterion: the redshift of
formation, $z_\mathrm{f}$, of a galaxy observed (as an ERO for instance) at a certain
redshift has to be consistent with the assumed cosmology. As will be
seen, this cosmological constraint rules out that ellipticals 
capable of being classified as EROs are older than about $7-8$ Gyr. 
\par
The second approach is to study how the $R-K$ colour evolves with
redshift. This can be done both by assuming a redshift of formation, $z_\mathrm{f}$,
and let a galaxy evolve in time, or by redshifting a galaxy of a fixed age.

\subsection{Overlap in the colour plane}
\label{distinguishingmethod}
The main aim of this work is to investigate whether broad band colour-colour
plots can be used to discriminate between the two main classes of galaxies
making up the ERO population: evolved ellipticals and dusty starbursts.
The colours of elliptical and starburst galaxies at
different redshifts may or may not overlap, and there could exist
redshift ranges in which the colour-colour regions of the two galaxy
populations can be distinctly defined. To investigate this, the first
step is to build up a grid of colours of ERO-classified galaxies. For a range of
redshifts, the colours of ellipticals and starbursts are computed. After removing
those that are neither consistent with the cosmology nor have $R-K
\geq 5$, two sets of colour-colour points are obtained, associated
with the EROs in each galaxy population. 
\par
The overlap that occurs in a colour-colour diagram between the colours
of the two galaxy populations at different redshifts is then
computed. The confusion, or overlap, between 
these two point sets are defined as follows. If $A$ designates the region
covered by the one set, and $B$ the region of the other set, then the
intersection of these two sets, $A\cap B$, is the region where the two sets
overlap. The measure of overlap is now taken as the area of the intersection
normalised by the area of the smallest region, i.e. 
\begin{equation}
    \mathrm{overlap} = \displaystyle\frac{A\cap B}{\min \{A, B\}} = \max \left\{\displaystyle\frac{A\cap B}{A}, \displaystyle\frac{A\cap B}{B}\right\}.
\end{equation}
In this way, an overlap measure of $1$ is obtained if one of the sets is
completely embedded in the other.
\par
To define the regions covered by the point sets, every colour-colour point is
considered as defining the centre of a circle having a fixed (small)
radius. If the point density is high enough, this will create a continuous
region. Two points are now regarded as overlapping if the distance between
them is less than twice the radii of the circles. This requires
that the point density is homogeneous, since it cannot be accepted that a higher
point density part of the colour-colour region will have a higher weighting in
the overlap computation. Hence, the two point sets are discretised into an 
evenly spaced grid, with a spacing of one circle radius. The colour-colour
points will be offset from their true loci by at most one half of a circle
radius. In the overlap computation, the circle radius is set to $0.05$
magnitudes, which will make two points more than $0.1$ magnitudes apart seen as
separate. The gridding will shift the colour-colour points by at most
$0.025$ magnitudes, which is well below the typical photometric errors.
\par
This procedure results in maps of overlap 
measures where one of the axes corresponds to the redshifts of the ellipticals, 
and the other to the redshifts of the starbursts (see Fig.~\ref{bestcombs}). From 
these maps, the redshift regions where the starburst EROs and the elliptical 
EROs can be separated from each other in a colour-colour diagram can be 
clearly seen. What is left is to explore these redshift regions to see
what colours the different galaxies have.

\section{Results and discussion}
\label{results_section}

\subsection{Colour evolution}

\subsubsection{The ellipticals}

All ellipticals have $R - K > 5$ at some redshift and the general trend of
the $R-K$ colour is to increase with redshift. The explanation for this can be
seen in the left panel of Fig.~\ref{spectra}: when the SED of an elliptical is
redshifted, it will be shifted to the right in the diagram, and except for the
youngest ellipticals, the $R$ filter will soon measure in the less luminous
region blue-ward of the $4000$ \AA\ break, while the $K$ filter measures
brighter and brighter parts of the SED. A higher metallicity reddens the SED
due to line-blanketing and backwarming in the stellar atmospheres.
A redder galaxy will require a lower redshift to make the galaxy extremely red, i.e.\ 
a lower $z_\mathrm{ERO}$ (as defined in Sect.~\ref{erosidentsec}). As a result of this, an
elliptical of higher metallicity can be encountered as an ERO over a larger redshift
range and hence it can also be older compared to an ERO elliptical of lower metallicity. 
\par
\begin{figure}
\centering
\resizebox{\hsize}{!}{\includegraphics{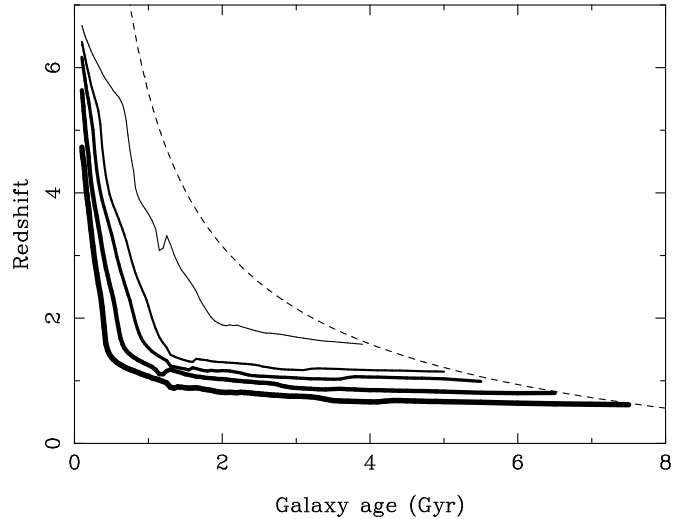}}
\caption{$z_\mathrm{ERO}$ vs.\ galaxy age for the
  ellipticals of Salpeter IMF, where $z_\mathrm{ERO}$ is defined as the lowest redshift where an
  elliptical can be an ERO. The metallicities ($0.02-2.5$ Z$_\odot$) of the models increase
  downwards/leftwards and the lines thicken accordingly. The
  \emph{dashed line} traces the cosmological redshift-age relation.}
\label{zeros}
\end{figure}

In Fig.~\ref{zeros} the $z_\mathrm{ERO}$s for the ellipticals are plotted
as a function of galaxy age. The dashed line shows the redshift-age
correspondence for the chosen cosmology: the area leftwards/downwards of this
line contains the redshift-age set consistent with the cosmology. The
five solid lines show the $z_\mathrm{ERO}$s for ellipticals of
different metallicities ($0.02-2.5$ Z$_\odot$), and the
thickness of the lines increases with increasing metallicity. Since
$z_\mathrm{ERO}$ is the lowest redshift needed to produce an ERO, the redshift
regime between the ``cosmology line'' and each $z_\mathrm{ERO}$ line shows the
redshifts where the EROs exist. It is clearly seen that the cosmology
requires the ellipticals to be younger than $7.5$ Gyr. Furthermore,
the lowest $z_\mathrm{ERO}$ is about $0.7$, a result which is
consistent with the observations of \citet{moustakas_et_al_2003}, who
find very few early-type EROs at redshifts below $\sim\!0.9$.
\par
\begin{figure*}
\centering
\resizebox{17cm}{!}{
  \includegraphics[width = 88mm]{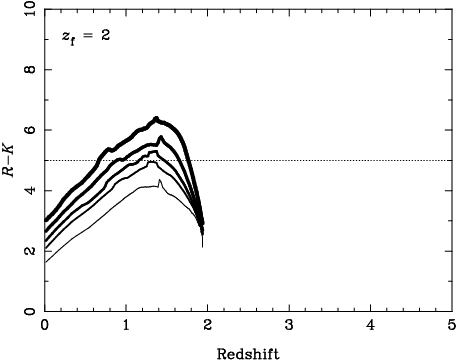}
  \hspace{3em}
  \includegraphics[width = 88mm]{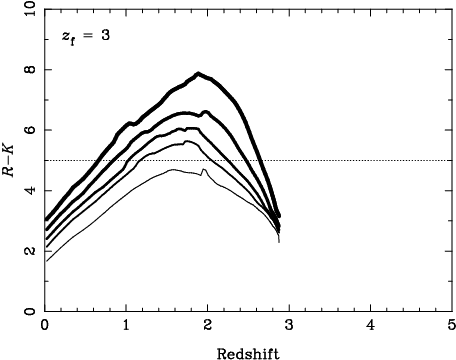}
}
\vspace{12pt} \
\resizebox{17cm}{!}{
  \includegraphics[width = 88mm]{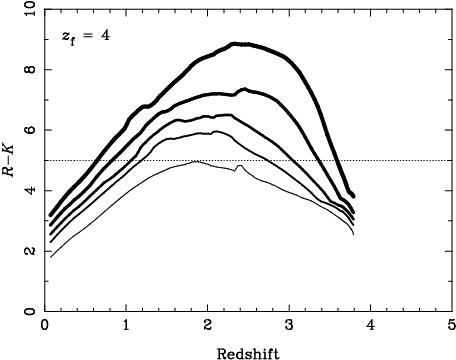}
  \hspace{3em}
  \includegraphics[width = 88mm]{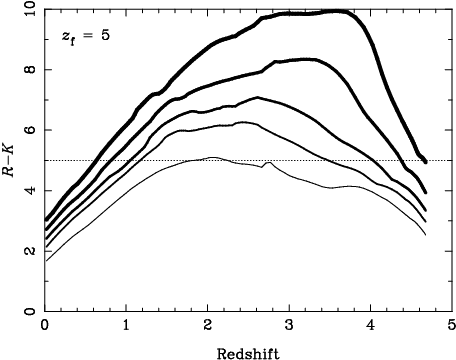}
}
\caption{Colour-redshift diagrams for the ellipticals of Salpeter IMF
  assuming formation redshifts from $2$ to $5$ using the full range of
  metallicities ($0.02-2.5$ Z$_\odot$). The thickness of the lines
  increases according to the metallicity.} 
\label{instRKzf}
\end{figure*}

The elliptical $R-K$ colour for different assumed redshifts of formation is shown
in Fig.~\ref{instRKzf}. The lowest metallicity, $Z = 0.02$ Z$_\odot$, only produces EROs in
a small redshift interval around $z = 2$ for the highest $z_\mathrm{f}$. For decreasing $z_\mathrm{f}$, the
limiting ERO-producing metallicity increases, and when $z_\mathrm{f} =
1$, not even the highest metallicity will yield red enough galaxies.

\subsubsection{The starbursts}

\begin{figure}
\centering
\resizebox{\hsize}{!}{\includegraphics{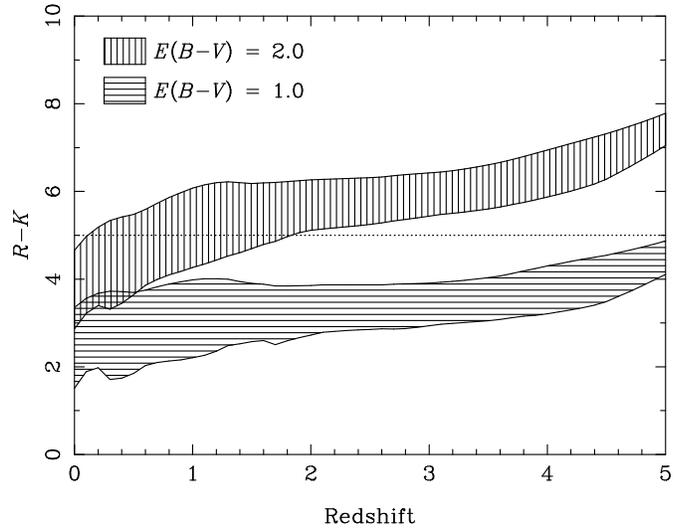}}
\caption{Colour-redshift diagram for the starbursts of Salpeter
  IMF. The vertically hatched region corresponds to $E(B-V)=2.0$, and
  the horizontally hatched region to $E(B-V)=1.0$. The full age and
  metallicity ranges are used; the more metal-rich and the older the
galaxy, the redder.} 
\label{contRKz}
\end{figure}

For starburst galaxies, the evolution of the $R-K$ colour with
redshift puts constraints on the amount of reddening needed to produce
EROs. This limiting amount of extinction will decrease as the metallicity
increases, since a more metal-rich galaxy will be redder.
The colour-redshift diagram in Fig.~\ref{contRKz} shows the regions of the
$R-K$ colour of the starbursts for colour excesses of $1.0$ and $2.0$. The
full ranges of galaxy age ($0-100$ Myr) and metallicity ($0.02-2.5$ Z$_\odot$)
are used. A colour excess of slightly more than $1$ is required to produce
EROs for $z\leq 5$. Excepting the very lowest redshifts, the starbursts can be observed as EROs
at all $z\leq 5$ when $E(B-V) = 2.0$.

\subsection{The colour plane properties}
\label{colourplane}

The following investigation of the colour-colour space, both as the overlap maps
and the colour-colour plots are concerned, does not depend on any
assumed formation redshifts. The galaxies are assumed to form
continuously, and only the cosmology sets the age limits for the
galaxies. At a certain redshift the considered galaxy models are those
that strictly fit into the current lifetime of the Universe. No
assumption is made of a redshift where the galaxy formation is started in the
Universe as a whole.

\subsubsection{The overlap in the colour-colour diagrams}
\label{filterconfigurationproperties}
Distinguishing between the elliptical and starburst models in
colour-colour diagrams is possible only if the colours of the galaxies
follow evolutionary tracks that do not overlap. As a first test,
colour-colour plots of all the $45$ colour-colour configurations that
can be formed from the filters $RIJHK$ are made for the whole
redshift grid (redshifts below $5$ in steps of $0.1$). Both 
Salpeter and Kroupa IMFs are used, and the colour excess of the starbursts
ranges up to 2.0 in steps of $0.1$. It 
is found that when considering this broad redshift range, all filter
configurations produce colour-colour plots where the region of the
starburst colours lie almost completely inside the colour region of
the ellipticals, hence they cannot be separated in this type of plots.
\par
\begin{figure*}
    \centering
    \mbox{
    \includegraphics{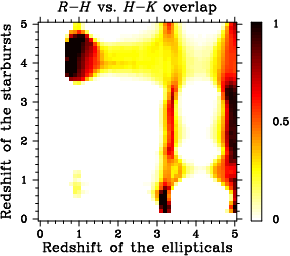}
    \hspace{2em}
    \includegraphics{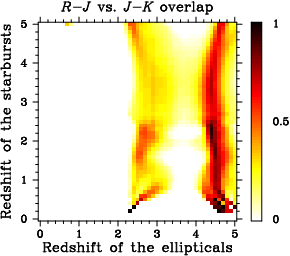}
    \hspace{2em}
    \includegraphics{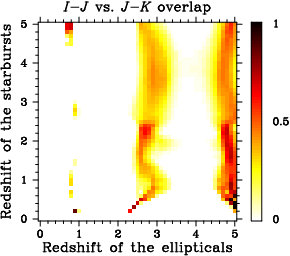}}
    \vspace{4pt} \ 
    \mbox{
    \includegraphics{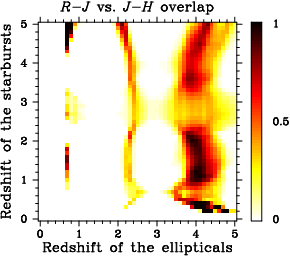}
    \hspace{2em}
    \includegraphics{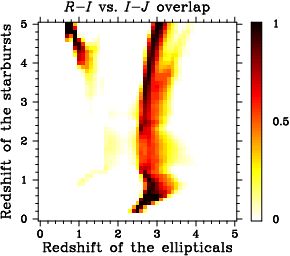}
    \hspace{2em}
    \includegraphics{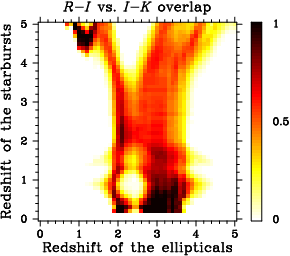}}
    \vspace{4pt} \ 
    \mbox{
    \includegraphics{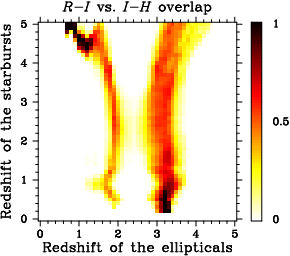}
    \hspace{2em}
    \includegraphics{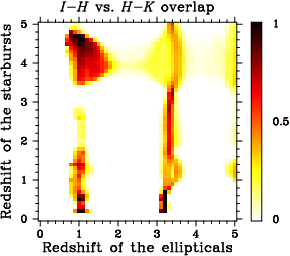}
    \hspace{2em}
    \includegraphics{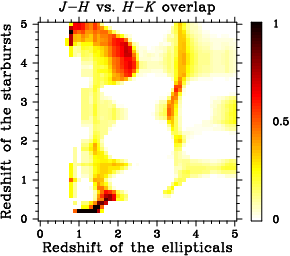}}
    \vspace{4pt} \ 
    \mbox{
    \includegraphics{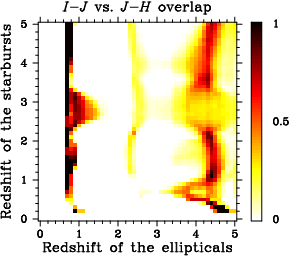}
    \hspace{2em}
    \includegraphics{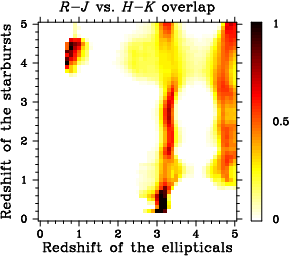}
    \hspace{2em}
    \includegraphics{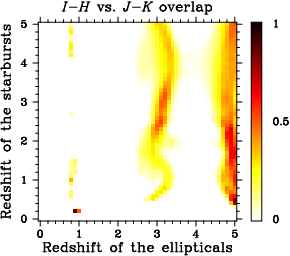}}
    \vspace{4pt} \ 
    \mbox{
    \includegraphics{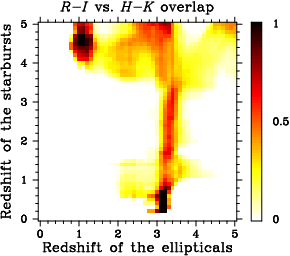}
    \hspace{2em}
    \includegraphics{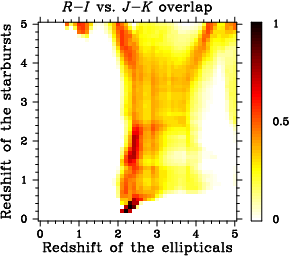}
    \hspace{2em}
    \includegraphics{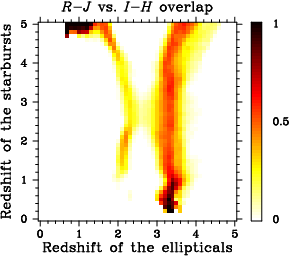}}
  \caption{The colour overlap maps for the filter configurations yielding the best 
  discriminator performance within each filter set. The brighter the area, the less
  overlap; white corresponds to a separation of the elliptical and
  starburst EROs by at least $0.1$ magnitudes in the corresponding
  colour-colour plot. The galaxy models have Salpeter and Kroupa IMFs, metallicities 
  from $0.02$ to $2.5$ Z$_\odot$, and for the starbursts, colour excesses up to
  $2.0$. Only models yielding EROs consistent with the cosmology are
  considered in computing the colour overlap.}
  \label{bestcombs}
\end{figure*}

The next step is to see in which redshift ranges the different configurations
produce colours where the galaxy populations are distinguishable, and the overlap
is computed using the procedure described in Sect.~\ref{distinguishingmethod}.
The model grid is again set up using Salpeter and Kroupa IMFs and
colour excesses of the starbursts up to $2$ in steps of $0.1$. The redshift
grid spacing is $0.1$. Overlap maps resulting from this procedure are shown in
Fig.~\ref{bestcombs}.  
\par 
A proper interpretation of these maps requires some remarks. (\textbf{i})~The overlap has
to be small over large continuous regions in the redshift space, starting from
$z=0$ and up to high redshifts. A colour configuration is of course only useful
as a discriminator at redshifts below the redshift where the colours
starts to overlap.
(\textbf{ii})~Since the fraction of EROs in a given galaxy population increases with
redshift, the region occupied by EROs in a colour-colour plot will
increase with redshift as well. At low redshifts, the colour region
covered by the starbursts can therefore be very small, which will make
any overlap with the ellipticals to appear more significant than what
is reasonable. Hence it is not enough to decide on a
redshift limit for a certain filter configuration from its overlap
map, but to determine the impact of the overlap at low redshifts, the
corresponding colour-colour diagram has to be
investigated. (\textbf{iii})~The ridges in the overlap maps often
represent cases where the colours of one galaxy population cross over
the colour region of the other. This causes the colour-colour plots
to exhibit very complicated structures which makes it difficult to
define a separation line.
\par
The configurations that can be formed from permutations of a set of three
bands (e.g.\ the triplet $\{RHK\}$ from which the filter configurations
$RHHK$\footnote{The notation ``$XYZW$'' will
  be used as an abbreviated form of ``$X-Y$ vs.\ $Z-W$''.}, $RKHK$ and $RHRK$
may be formed) have a ``pivot band'' against which 
the two other magnitudes are compared, e.g.\ the $K$ band in the $RKHK$
configuration. The configurations formed from such a triplet are found to have
very similar overlap features, although small differences can be seen. For all
the triplets, the configurations where the pivot wavelength band lies between
the other two bands, are \emph{always} the best ones; for instance
$RHHK$ in the triplet $\{RHK\}$. It should be noted that this result is
consistently observed throughout the course of this work. The worst performance
is in most cases exhibited by the configurations using the shorter wavelength
band as pivot band, although some (small) changes can be seen if higher
extinctions are included. 
\par
The configurations formed from quadruples (configurations made with four
different bands, e.g.\ the quadruple $\{RJHK\}$ from which the
configurations $RJHK$, $RHJK$ and $RKJH$ may
be formed) tend to behave very differently and cannot be regarded as forming
distinct groups having similar overlap features. Since they are based on four
regions on the SED, they are more sensitive to differences between the
SEDs. However, in the overlap maps this advantage over the triplets manifests
itself only in that the average performance of the quadruples is better
than the triplets, but the best triplet has still a higher redshift
limit than the best quadruple configuration. Furthermore, the \emph{only}
quadruple that actually performs better than the four triplets it can be
broken into is the $\{IJHK\}$ quadruple. This combination of filters appears
unfortunate; the triplets $\{IHK\}$, $\{JHK\}$ and 
$\{IJH\}$ are the three worst of the triplets, and the $\{IJK\}$
as well as $\{IJHK\}$ itself both yield complicated colour-colour
plots. For the other quadruples, there is always a configuration of three
filters that performs at least as well as its ``parent'' quadruple.
\par
The relative discriminating performance of the three configurations within each
filter set are shown in Table \ref{colourplanetable}. Figure
\ref{bestcombs} shows the filter configuration yielding the best 
overlap properties within a given filter set. The colour coding means the
brighter the area, the less overlap. The $RHHK$ configuration has one of the highest
redshift limits, and can discriminate between the galaxy populations up to
redshift $2.9$. The small overlap between ellipticals and starbursts at $z\sim
1$ for this configuration does not pose a problem. This will be discussed further in the
next section.

\begin{table*}
\centering
\caption{This table shows (\textbf{i}) The relative performance of the filter configurations within
   each filter set. This is based on the ability to discriminate
   between the colours of the modelled ellipticals and starbursts for
   $E(B-V) \leq 2$. (\textbf{ii}) The separation line corresponding to the configuration that
  has the best discriminating performance within each filter set. (\textbf{iii}) The redshift limit beyond
  which the colour regions become entangled, $z_\mathrm{max}$. The
   separation line is valid up to this redshift limit. Separation
   lines are not defined for the filter configurations having a
   redshift limit less than about one. (\textbf{iv}) The faintest ERO $K$ magnitude
   of $L^*$ ellipticals and starbursts at $z_\mathrm{max}$.} 
\label{colourplanetable}
\begin{tabular}{lllll@{ $ > $ }r@{ }l@{ }l@{ }llll}
\hline
\hline
Filter set & \multicolumn{3}{l}{Configuration performance} &
\multicolumn{5}{l}{Separating line: colours of} & $z_\mathrm{max}$ &
\multicolumn{2}{c}{$K(L_\mathrm{bol} = L^*)$ at $z_\mathrm{max}$} \\
     & Best & $\longrightarrow$ & Worst & \multicolumn{5}{l}{ERO ellipticals at $z \leq
  z_\mathrm{max}$} & & Ellipticals & Starbursts \\
\hline
$\{RHK\}$  & $RHHK$ & $RKHK$ & $RHRK$ & $(R-H)$ & $ 3.69$ & $(H-K)$ & $+$ & $0.60$               & $2.9$ & $20.7$ & $26.1$ \\
$\{RJK\}$  & $RJJK$ & $RKJK$ & $RJRK$ & $(R-J)$ & $ 1.99$ & $(J-K)$ & $-$ & $1.18^\mathrm{a}$    & $2.2$ & $20.7$ & $25.4$ \\
$\{IJK\}$  & $IJJK$ & $IKJK$ & $IJIK$ & $(I-J)$ & $ 1.21$ & $(J-K)$ & $-$ & $0.64^\mathrm{a,b}$  & $2.2$ & $20.7$ & $25.4$ \\
$\{RJH\}$  & $RJJH$ & $RHJH$ & $RJRH$ & $(R-J)$ & $ 1.77$ & $(J-H)$ & $+$ & $1.10$               & $2.1$ & $20.7$ & $25.3$ \\
$\{RIJ\}$  & $RIIJ$ & $RJIJ$ & $RIRJ$ & $(R-I)$ & $-0.38$ & $(I-J)$ & $+$ & $2.05$               & $1.6$ & $20.7$ & $24.8$ \\
$\{RIK\}$  & $RIIK$ & $RIRK$ & $RKIK$ & $(R-I)$ & $ 0.22$ & $(I-K)$ & $+$ & $0.39$               & $1.4$ & $20.6$ & $24.6$ \\
$\{RIH\}$  & $RIIH$ & $RHIH$ & $RIRH$ & $(R-I)$ & $ 0.25$ & $(I-H)$ & $+$ & $0.56$               & $1.2$ & $20.5$ & $24.4$ \\ 
$\{IHK\}$  & $IHHK$ & $IKHK$ & $IHIK$ & \multicolumn{5}{l}{}                                     &       &        &        \\
$\{JHK\}$  & $JHHK$ & $JKHK$ & $JHJK$ & \multicolumn{5}{l}{}                                     &       &        &        \\
$\{IJH\}$  & $IJJH$ & $IHJH$ & $IJIH$ & \multicolumn{5}{l}{}                                     &       &        &        \\
\hline
$\{RJHK\}$ & $RJHK$ & $RHJK$ & $RKJH$ & $(R-J)$ & $ 4.59$ & $(H-K)$ & $-$ & $1.44^\mathrm{a}$    & $2.6$ & $20.7$ & $25.8$ \\
$\{IJHK\}$ & $IHJK$ & $IJHK$ & $IKJH$ & $(I-H)$ & $ 1.78$ & $(J-K)$ & $-$ & $0.72^\mathrm{a,b}$  & $2.4$ & $20.7$ & $25.6$ \\
$\{RIHK\}$ & $RIHK$ & $RHIK$ & $RKIH$ & $(R-I)$ & $ 1.80$ & $(H-K)$ & $-$ & $0.59^\mathrm{a}$    & $2.0$ & $20.7$ & $25.2$ \\ 
$\{RIJK\}$ & $RIJK$ & $RJIK$ & $RKIJ$ & $(R-I)$ & $ 0.78$ & $(J-K)$ & $-$ & $0.35^\mathrm{a}$    & $1.8$ & $20.7$ & $25.0$ \\ 
$\{RIJH\}$ & $RJIH$ & $RIJH$ & $RHIJ$ & $(R-J)$ & $ 0.65$ & $(I-H)$ & $+$ & $1.20$               & $1.8$ & $20.7$ & $25.0$ \\
\hline 
\multicolumn{12}{l}{}\\
\multicolumn{12}{l}{$^\mathrm{a}$ Line parallel with reddening vector.} \\
\multicolumn{12}{l}{$^\mathrm{b}$ Line runs partly through the
  colour-colour regions due to complex region shapes.} \\
\end{tabular}
\end{table*}

\begin{table*}
\centering
\caption{The consistency of the classifications made using different filter configurations 
are checked against classifications of samples of EROs in the literature. The consistency 
is grouped according to classification method: (\textbf{i}) the photometric classifications by 
\citet{smail_et_al_2002}; (\textbf{ii}) the spectroscopic classifications by 
\citet{graham_dey_1996}, \citet{spinrad_et_al_1997}, \citet{soifer_et_al_1999},
\citet{afonso_et_al_2001}, \citet{smith_et_al_2001},
\citet{frayer_et_al_2003} and \citet{saracco_et_al_2003}; and (\textbf{iii}) the
morphologic classifications by \citet{franceschini_et_al_1998},
\citet{stiavelli_et_al_1999}, \citet{smith_et_al_2002a} and 
\citet{van_dokkum_et_al_2003}. The consistency is given both as a
\emph{colour consistency}, i.e. the consistency with the separation
lines given in Table \ref{colourplanetable}, and a \emph{region
  consistency}, i.e. if the sources fall in the proper colour-colour
regions. The consistency is given as the number of sources on the
correct side of the separation line (\emph{or} in the correct colour
region) over the respective number of sources available.} 
\label{consistencytable}
\begin{tabular}{lll|r@{ $/$ }lr@{ $/$ }l|r@{ $/$ }lr@{ $/$ }l|r@{ $/$ }lr@{ $/$ }l}
\hline
\hline
Filter set & Configuration & $z_\mathrm{max}$ & \multicolumn{12}{c}{Colour (region) consistency for different 
                                                                   classification methods} \\
\hline
           &               &                  & \multicolumn{4}{c|}{Smail et al. 2002 sample}
                                              & \multicolumn{4}{c|}{Spectroscopically classified} 
                                              & \multicolumn{4}{c}{Morphologically classified} \\
           &               &                  & \multicolumn{2}{l}{Evolved} & \multicolumn{2}{l|}{Dusty}   
                                              & \multicolumn{2}{l}{Evolved} & \multicolumn{2}{l|}{Dusty}  
                                              & \multicolumn{2}{l}{Evolved} & \multicolumn{2}{l}{Dusty} \\
\hline
$\{RHK\}$  & $RHHK$ & $2.9$ & $6(6)$ & $ 8$ & $11(11)$ & $15$ & $2(2)$ & $2$ & $1(1)$ & $1$ & $1(1)$ & $1$ & \multicolumn{2}{c}{} \\
$\{RJK\}$  & $RJJK$ & $2.2$ & $9(8)$ & $ 9$ & $3(1^\mathrm{a})$ & $ 9$ & $4(4)$ & $4$ & $2(2)$ & $3$ & $3(3)$ & $4$ & $0(0)$ & $10$ \\
$\{IJK\}$  & $IJJK$ & $2.2$ & $7(4)$ & $ 7$ & $4(3)$   & $10$ & $3(2)$ & $3$ & $0(0^\mathrm{a})$ & $3$ & $20(11)$ & $21$ & $1(1^\mathrm{a})$ & $9$ \\
$\{RJH\}$  & $RJJH$ & $2.1$ & $4(2)$ & $ 5$ & $1(1)$   & $ 9$ & $2(2)$ & $2$ & $0(0)$ & $1$ & $0(0)$ & $1$ & \multicolumn{2}{c}{} \\
$\{RIJ\}$  & $RIIJ$ & $1.6$ & $5(3)$ & $ 7$ & $5(3)$   & $ 9$ & $3(1)$ & $3$ & $1(0)$ & $2$ & $4(2)$ & $4$ & $0(0)$ & $9$ \\
$\{RIK\}$  & $RIIK$ & $1.4$ & $3(2)$ & $10$ & $11(10)$ & $17$ & $1(1)$ & $3$ & $1(0^\mathrm{a})$ & $2$ & $4(4)$ & $8$ & $9(7^\mathrm{a})$ & $20$ \\
$\{RIH\}$  & $RIIH$ & $1.2$ & $2(0)$ & $ 8$ & $9(6)$   & $15$ & $1(0)$ & $1$ & $0(0)$ & $1$ & $0(0)$ & $1$ & \multicolumn{2}{c}{} \\ 
\hline
$\{RJHK\}$ & $RJHK$ & $2.6$ & $5(5)$ & $5$ & $7(3^\mathrm{a})$ & $ 9$ & $2(2)$ & $2$ & $1(0^\mathrm{a})$ & $1$ & $1(1)$ & $1$ & \multicolumn{2}{c}{} \\
$\{IJHK\}$ & $IHJK$ & $2.4$ & $5(4)$ & $5$ & $5(3)$    & $10$ & $1(1)$ & $1$ & $1(1)$ & $2$ & $17(11)$ & $18$ & \multicolumn{2}{c}{} \\
$\{RIHK\}$ & $RIHK$ & $2.0$ & $5(5)$ & $8$ & $11(7^\mathrm{a})$ & $15$ & $1(1)$ & $1$ & $1(0^\mathrm{a})$ & $1$ & $0(0)$ & $1$ & \multicolumn{2}{c}{} \\ 
$\{RIJK\}$ & $RIJK$ & $1.8$ & $5(3)$ & $7$ & $6(3)$    & $ 9$ & $3(3)$ & $3$ & $2(0^\mathrm{a})$ & $2$ & $2(2)$ & $4$ & $4(3^\mathrm{a})$ & $9$ \\ 
$\{RIJH\}$ & $RJIH$ & $1.8$ & $4(2)$ & $5$ & $2(0)$    & $ 9$ & $1(1)$ & $1$ & $0(0)$ & $1$ & $0(0)$ & $1$ & \multicolumn{2}{c}{} \\
\hline 
\multicolumn{15}{l}{}\\
\multicolumn{15}{l}{$^\mathrm{a}$ The region consistency can be improved by using a higher extinction.} \\
\end{tabular}
\end{table*}

\subsubsection{Non-overlapping colour regions}
\label{nonoverlapping}

Having found the redshift regimes where the best colour-colour plots exhibit
distinct elliptical and starburst colour regions as described in
Sect.~\ref{filterconfigurationproperties}, an attempt is now
made to define lines that separate the two regions. A filter configuration can 
be considered as having an improved performance as a discriminator if
the separation line is parallel with the reddening vector, since
an increase in extinction will not cause the starbursts to cross such a
separation line. However, as a higher extinction also will enable lower-redshift
starbursts to become EROs, it is possible that these ``new''
candidates will intrude on the regions of the ellipticals.
\par
A separation line between the colour regions
of ellipticals and starbursts are defined for the configuration
having the best performance within each filter set, see Table
\ref{colourplanetable}. For a given colour-colour configuration, it
is in many cases possible to define a whole set of lines separating
the two regions; in these cases a representative line is
presented. The separation line is valid for $z \leq z_\mathrm{max}$,
where $z_\mathrm{max}$ is the redshift limit beyond which
the colours will be confused. A separation line is not defined for the
filter sets that have redshift limits below about one, as non-overlap
at such low redshifts are due to a lack of EROs among the
ellipticals and/or the starbursts.  
\par
To check the performance of the different filter configurations,
comparisons are made with existing classifications in the
literature. This is shown in Table \ref{consistencytable}. The
consistency of the classifications made here is checked according to
classification method: photometric, spectroscopic or
morphologic. The photometric consistency check is made against the
data in \citet[][ S02 hereafter]{smail_et_al_2002}. S02 have obtained
a sample of EROs from deep $RIzJHK$ photometry and have also
classified the sources by template fitting. The sample is thus divided
into two classes where the photometry is consistent with ``dusty'' or
``evolved'' galaxies.  The spectrocopically
classified set is obtained from \citet{graham_dey_1996},
\citet{spinrad_et_al_1997}, \citet{soifer_et_al_1999},
\citet{afonso_et_al_2001}, \citet{smith_et_al_2001},
\citet{frayer_et_al_2003} and \citet{saracco_et_al_2003}.
The morphologically classified EROs are obtained from
\citet{franceschini_et_al_1998}, \citet{stiavelli_et_al_1999},
\citet{smith_et_al_2002a} and
\citet{van_dokkum_et_al_2003}. \citet{smith_et_al_2002a} use the
classification ``Irregular'' and ``Compact'', which, for the
consistency check, is simply interpreted as ``starburst'' and
``elliptical'' (or ``dusty'' and ``evolved''). This might be too
simplistic which would explain the apparently large discrepancy for
the morphologically classified starbursts, since the
\citet{smith_et_al_2002a} sample comprises the majority of these
sources.
\par
\begin{figure}
\centering
\resizebox{\hsize}{!}{\includegraphics{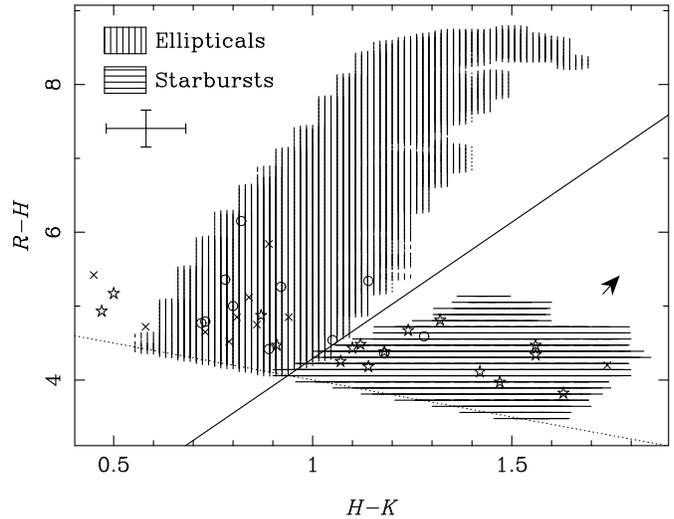}}
\caption{$R-H$ vs.\ $H-K$ regions for the Salpeter and Kroupa IMF models
  for $z \leq 2.9$ having extinctions $E(B-V) \leq 2$, plotted
  together with classified EROs (references in text and in Table
  \ref{consistencytable}). \emph{Stars} denote dusty galaxies,
  \emph{circles} evolved galaxies, and \emph{crosses} denote the
  unclassified detections by \citet{smail_et_al_2002}. The arrow
  indicates how the starburst region shifts if the colour excess is
  increased by $0.1$. The error bar shows the typical photometric error of the
  colours of the EROs.} 
\label{RHHK}
\end{figure}

The $RHHK$ configuration is the best one as it concerns
redshift, and it also agrees well with existing classifications. Figure \ref{RHHK} shows the
$RHHK$ colour diagram for starbursts and ellipticals up to redshift
$2.9$. This filter configuration agrees with all the spectroscopically
and morphologically classified sources -- although the number of sources with $H$
band data is limited -- and with the classifications made by S02
for $11$ of $15$ dusty and $6$ of $8$ evolved galaxies that have data in these
three bands. Furthermore, one of the $11$ unclassified S02 sources is found to
have the colours of starbursts and $8$ (possibly $10$) resemble
ellipticals. The sources that end up inside the ``wrong'' colour region can be
explained as lower-redshift starbursts having colour excesses between $2$ and
$3$ and ellipticals at redshifts higher than $2.9$, respectively. This
combination has the property of being a minimal choice of filters, since it
only requires one extra filter except for the ERO criterion filters. Figure
\ref{compare} shows how the $R-K$ colour can select the extremely red
elliptical and starburst galaxies, and how they can be distinguished between
in the $H$ band. 

\par
The $RJJK$ configuration has a smaller redshift coverage, but is also a
minimal choice of filters. However, this does not agree with the existing
classifications very well, in particular the consistency for the
starbursts is bad. In this configuration, a majority of the sources have colours
that are consistent with the colours of the model ellipticals. The mismatch
also occurs for the $\{IJK\}$ configurations, where
additionally the complicated structure of the colour-colour regions makes it
difficult to define a clear separation line. The configurations in the
$\{IHK\}$, $\{JHK\}$ and $\{IJH\}$ triplets have all very
high overlap at low redshifts, which can be attributed to the problem lined
out in remark (\textbf{ii}) in Sect.~\ref{filterconfigurationproperties}
above. However, as these high overlap regions extends over such considerable
redshift space, these filter configurations must be regarded as having a poor
discriminating ability. 
\par
The best quadruple configuration is $RJHK$. The separation line in this case
agrees well with earlier classifications, but a majority of the ``dusty''
S02 sources fall slightly outside the starburst
region. A heavier extinction could justify this, though.  
\par
A problem with this kind of method to classify EROs is
indicated by the last two columns in Table \ref{colourplanetable}. These show the
$K$ magnitude of the faintest $L^*$ ($2\times10^{10}$ L$_\odot$) ERO at the limiting redshift for
each colour plane. In general, an $L^*$ starburst ERO is $4-6$ 
magnitudes fainter in the $K$ band than an elliptical at
$z_\mathrm{max}$. This points to a severe difficulty for studies of
EROs aimed at disentangling this population: when a study is deep
enough to detect ``ordinary'' starburst galaxies, ellipticals at
redshifts far beyond $z_\mathrm{max}$ will also be detected. Without
redshift information, these high-redshift ellipticals will be included
in the sample, and the result will be colour planes with a large
amount of confusion. On the other hand, a more shallow survey will
tend to detect a majority of ellipticals, which can mislead the
conclusions.  
\par
In its extension, this problem can affect conclusions drawn from
studies aimed at determining the fractions of different galaxy types
among the EROs using \textit{any} of the standard methods.
Considering these differences in magnitude of the two
populations, it seems plausible that the conclusion that ellipticals
dominate the ERO population might just be an effect of this
bias. However, a thorough study of the luminosity functions of 
the galaxy populations constituting the ERO class is required to
resolve these implications. The widely different characteristics of
the two main candidate populations will presumably make studies aimed at
determining the relative fractions of galaxy types among the EROs very sensitive
to biases.

\subsubsection{Redder EROs}
The effect of applying higher extinctions is that a greater number of
lower-redshift starbursts will be included in the ERO population, some of
which will have colours similar to the ellipticals. As mentioned before, the 
starbursts which are on the correct side of a separation line that is
parallel to the reddening vector will not cross the line if the extinction
is increased. However, the new ERO candidates among the low-redshift
starbursts which previously had $R-K < 5$ can, as they are reddened further,
appear ``from below'' and on the wrong side of the separation line, thus
ending up in the region of the ellipticals. A conclusion from this, is that
the redshift plays a major part in determining which side of the separation line that the
starbursts populate. In fact, this appears to be a general effect of the
redshift, since if higher-redshift ellipticals are included, these will move
towards the separation line as the redshift is increased, and eventually cross
the line. 
\par
\begin{figure}
\centering
\resizebox{\hsize}{!}{\includegraphics{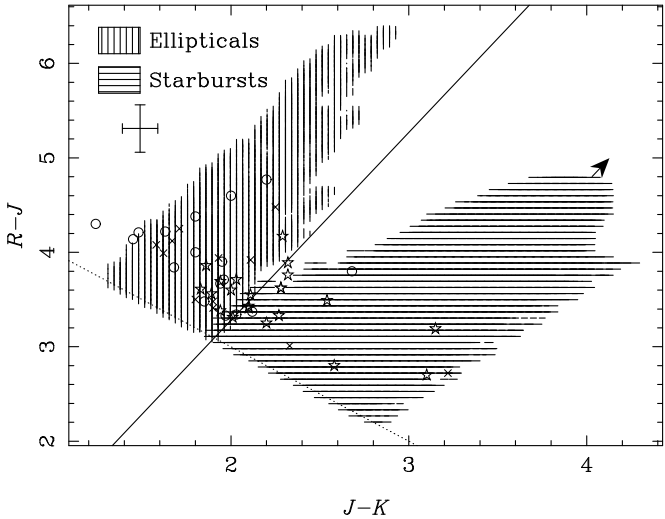}}
\caption{$R-J$ vs.\ $J-K$ regions for the Salpeter and Kroupa IMF
  models for $z \leq 2$ having extinctions $E(B-V) \leq 3$, plotted
  together with classified EROs (references in text and in Table
  \ref{consistencytable}). \emph{Stars} denote dusty galaxies,
  \emph{circles} evolved galaxies, and \emph{crosses} denote the
  unclassified detections by \citet{smail_et_al_2002}. The arrow indicates how
  the starburst region shifts if the colour excess is increased by
  $0.1$. The error bar shows the typical photometric error of the colours of the
  EROs.} 
\label{RJJK}
\end{figure}

A higher extinction does in general affect the triplet configurations more than the quadruples. For
$E(B-V)\leq 3$, the most stable triplet configuration is $RJJK$, which is
valid for $z\leq 2$, although it has a considerable overlap between starbursts
at $z\leq 0.5$ and ellipticals at $z\sim 0.8$. Accepting this overlap, the
proper separation line is $(R-J)=1.99(J-K)-0.70$; see Fig.~\ref{RJJK}. The best quadruple
configuration when $E(B-V)\leq 3$ is $RJHK$, in which the ellipticals have colours
consistent with $(R-J)> 0.90(I-K)-0.28$ for $z\leq 1.8$. Note that these two
separation lines are parallel with the reddening vector. 

\par
The increased overlap at low redshifts caused by high extinction starbursts
can be remedied by using a more strict ERO criterion, such as $R-K>6$, which
will disentangle the colours by simply de-selecting galaxies, in particular
those at $z \la 1$, cf.\ Figs. \ref{instRKzf} and \ref{contRKz}. Hence, a stricter
criterion will stabilise the discriminating properties of the colour planes, and
also in general increase their redshift limits.
When the criterion $R-K>6$ is used, the two configurations that have the best
redshift coverage are the $RJJK$ (ellipticals have $(R-J) > 1.99 (J-K) -0.94$
for $z\leq 2.1$) and the $IJJK$ ($(I-J) > 1.21 (J-K) -0.30$ for $z 
\leq 2.2$). Also in these cases, the separation lines are parallel to the
reddening vector. The latter configuration will require an ERO selection criterion,
though. 
\par 
The $RHHK$ turns out not to be very stable for an increase of the
starburst extinction when using the standard ERO criterion ($R-K>5$), as
lower-redshift starbursts of high extinction are selected and fall in
the colour region of the ellipticals. Using the stricter ERO
criterion $R-K>6$, $RHHK$ is the best configuration for $E(B-V)\leq 2$ and
is now valid for $z \leq 3.2$. High-redshift ERO ellipticals now forces the
separation line to the right in Fig.~\ref{RHHK}; these are consistent with $(R-H) >
4.00(H-K)-0.15$. Pushing the colour excess to higher values shows that $RHHK$
can be used for $E(B-V)\leq 2.4$ when $z\leq 3$. For this high
extinction, the position of the separation line is dictated by the starbursts,
and is now moved slightly to the left, to $(R-H) > 3.69 (H-K) + 0.77$.
\par
Due to a possible shortcoming of existing spectral evolution
synthesis programmes, the modelled SEDs might in fact be too blue. As pointed out by
\citet{zackrisson_et_al_2001}, the pre-main sequence evolution of
stars will have a non-negligible effect on the SED during the first
tens of Myr for a continuous star formation, a fact which is usually
overlooked in existing programmes, such as P\'EGASE.2, among others. 
However, as this phase reddens the SED, this is not a serious
problem for this study, because (\textbf{i}) in colour-colour
diagrams, such as Figs. \ref{RHHK} and \ref{RJJK}, the starbursts will
move up and to the right, and hence away from the loci of the
ellipticals, and (\textbf{ii}) since this is a transient effect, the
younger starbursts are moved closer to the older starbursts in
Fig.~\ref{contRKz}, which actually narrows the area covered, hence
working to decrease the overlap.

\section{Summary and conclusions}
\label{summary_section}
In this paper the colour-colour properties of EROs are investigated. Two types
of galaxies, starbursts and ellipticals, have been modelled using the spectral
synthesis evolution program P\'EGASE.2. The ellipticals are modelled as dust-free
galaxies, passively evolving after an instantaneous burst of star formation.
The starbursts are modelled as having a constant star formation rate for 100
Myr, with dust extinction of various degrees following the Calzetti et al. (2000)
obscuration law. \citet{salpeter1955} and \citet[][ equation (6)]{kroupa_2001}
IMFs are used and the metallicities of the galaxies were evolved consistently
with age, using initial metallicities from $0.02$ up to $2.5$ Z$_\odot$. The colour
properties in the 45 colour planes formed from the $RIJHK$ bands are
explored up to redshift $5$ in a $H_0 = 70$ km$\,$s$^{-1}\,$Mpc$^{-1}$,
$\Omega_\mathrm{m} = 0.3$ and $\Omega_\Lambda = 0.7$ cosmology. 
\par
The following conclusions can be drawn.
\begin{enumerate}
\item Except for the very lowest-metallicity cases, the ellipticals
  will be EROs at any redshift beyond $\sim\!0.6$ in the adopted
  cosmology. Furthermore, the stellar population has to be formed before $z\sim 1$
  and to be consistent with cosmology, an ERO-classified elliptical
  has to be younger than $7-8$ Gyr.
\item The starburst galaxies can be detected as EROs at an extinction
  of $E(B-V) > 1$ at the very highest redshifts. When $E(B-V)= 2$ the
  starbursts have $R-K > 5$ at redshifts above $\sim\!0.1$. 
\item The $30$ colour-colour combinations that can be formed from three of the
  $RIJHK$ filters exhibit the feature that permutations of the filters
  yield very similar overlap maps. Among the configurations formed
  from these filter triplets, the ones having the centre waveband as pivot
  band (i.e.\ the $H$ band in the $R-H$ vs. $H-K$ configuration) have
  the best performance as discriminators. 
\item\label{item4} Of special interest is the $R-H$ vs.\ $H-K$ configuration.
  This provides a minimal choice of filters, as it contains the
  fundamental ERO criterion, and has a very low degree of overlap
  between the galaxy populations up to redshifts of $2.9$ for $E(B-V)
  \leq 2$. Using the alternative ERO criterion $R-K > 6$, the $R-H$ vs.\ $H-K$
  configuration works for $E(B-V) \leq 2.4$ and $z \leq 3$.
\end{enumerate}
\par
The results in Sect. \ref{colourplane}, of which item \ref{item4} is a subset, 
represent the limit of what is achievable with this method of photometric
classifications of EROs. Since the chosen star formation histories are
extremes (passive evolution versus constant star formation),
the separation of the colours of the modelled galaxies is maximised. A
galaxy whose star formation history have contributions from both modes
of star formation will be more difficult to classify \citep[c.f.][]{pierini_et_al_2003}. This photometric
method is useful for galaxies clearly dominated by either star formation
mode, but to securely establish the nature of an ERO other methods
have to be applied as well.

\acknowledgement{The authors acknowledge financial support from
  Vetenskapsr\aa det (VR; the Swedish Natural Science
Council) for this project. S.\ Bergstr\"om is financed through a VR Graduate
  Student Grant (doktorandtj\"anst).}

\bibliography{h4513}

\begin{thebibliography}{46}
\expandafter\ifx\csname natexlab\endcsname\relax\def\natexlab#1{#1}\fi

\bibitem[{{Afonso} {et~al.}(2001){Afonso}, {Mobasher}, {Chan}, \&
  {Cram}}]{afonso_et_al_2001}
{Afonso}, J., {Mobasher}, B., {Chan}, B., \& {Cram}, L. 2001, \apjl, 559, L101

\bibitem[{Bessell(1990)}]{bessell1990}
Bessell, M.~S. 1990, \pasp, 102, 1181

\bibitem[{Bessell \& Brett(1988)}]{bb1988}
Bessell, M.~S. \& Brett, J.~M. 1988, \pasp, 100, 1134

\bibitem[{Bruzual \& Charlot(1991)}]{bruzual_charlot_1991}
Bruzual, G.~A. \& Charlot, S. 1991, \apj, 367, 126

\bibitem[{Bruzual \& Charlot(1993)}]{bruzual_charlot_1993}
Bruzual, G.~A. \& Charlot, S. 1993, \apj, 405, 538

\bibitem[{Calzetti {et~al.}(2000)Calzetti, Armus, Bohlin,
  {et~al.}}]{calzetti2000}
Calzetti, D., Armus, L., Bohlin, R.~C., {et~al.} 2000, \apj, 533, 682 (C00)

\bibitem[{{Cimatti}(2001)}]{cimatti2001}
{Cimatti}, A. 2001, in Deep Fields, 81, astro-ph/0012057

\bibitem[{{Cimatti} {et~al.}(1998){Cimatti}, {Andreani}, {Rottgering}, \&
  {Tilanus}}]{cimatti_et_al_1998}
{Cimatti}, A., {Andreani}, P., {Rottgering}, H., \& {Tilanus}, R. 1998, \nat,
  392, 895

\bibitem[{Cimatti {et~al.}(1999)Cimatti, Daddi, di~Serego~Alighieri,
  {et~al.}}]{cimatti1999}
Cimatti, A., Daddi, E., di~Serego~Alighieri, S., {et~al.} 1999, \aap, 352, L45

\bibitem[{Cimatti {et~al.}(2002)Cimatti, Daddi, Mignoli,
  {et~al.}}]{cimatti_et_al_2002}
Cimatti, A., Daddi, E., Mignoli, M., {et~al.} 2002, \aap, 381, L68

\bibitem[{{Daddi} {et~al.}(2002){Daddi}, {Cimatti}, {Broadhurst}, {Renzini},
  {Zamorani}, {Mignoli}, {Saracco}, {Fontana}, {Pozzetti}, {Poli}, {Cristiani},
  {D'Odorico}, {Giallongo}, {Gilmozzi}, \& {Menci}}]{daddi_et_al_2002}
{Daddi}, E., {Cimatti}, A., {Broadhurst}, T., {et~al.} 2002, \aap, 384, L1

\bibitem[{Daddi {et~al.}(2000)Daddi, Cimatti, Pozzetti, Hoekstra,
  R{\"o}ttgering, Renzini, Zamorani, \& Mannucci}]{daddi_et_al_2000}
Daddi, E., Cimatti, A., Pozzetti, L., {et~al.} 2000, \aap, 361, 535

\bibitem[{{Dey} {et~al.}(1999){Dey}, {Graham}, {Ivison}, {Smail}, {Wright}, \&
  {Liu}}]{dey_et_al_1999}
{Dey}, A., {Graham}, J.~R., {Ivison}, R.~J., {et~al.} 1999, \apj, 519, 610

\bibitem[{{Eggen} {et~al.}(1962){Eggen}, {Lynden-Bell}, \&
  {Sandage}}]{eggen_lynden-bell_sandage_1962}
{Eggen}, O.~J., {Lynden-Bell}, D., \& {Sandage}, A.~R. 1962, \apj, 136, 748

\bibitem[{Elston {et~al.}(1988)Elston, Rieke, \& Rieke}]{err1988}
Elston, R., Rieke, G.~H., \& Rieke, M. 1988, \apjl, 331, L77

\bibitem[{Fioc \& Rocca-Volmerange(1997)}]{frv1997}
Fioc, M. \& Rocca-Volmerange, B. 1997, \aap, 326, 950

\bibitem[{Fioc \& Rocca-Volmerange(1999)}]{pegase}
Fioc, M. \& Rocca-Volmerange, B. 1999, astro-ph/9912179

\bibitem[{Franceschini {et~al.}(1998)Franceschini, Silva, Fasano, Granato,
  Bressan, {et~al.}}]{franceschini_et_al_1998}
Franceschini, A., Silva, L., Fasano, G., {et~al.} 1998, \apj, 506, 600

\bibitem[{Frayer {et~al.}(2003)Frayer, Armus, Scoville, Blain, Reddy,
  {et~al.}}]{frayer_et_al_2003}
Frayer, D.~T., Armus, L., Scoville, N.~Z., {et~al.} 2003, \aj, 126, 73

\bibitem[{Graham \& Dey(1996)}]{graham_dey_1996}
Graham, J.~R. \& Dey, A. 1996, \apj, 471, 720

\bibitem[{{Kauffmann} {et~al.}(1993){Kauffmann}, {White}, \&
  {Guiderdoni}}]{kauffmann_white_guiderdoni_1993}
{Kauffmann}, G., {White}, S.~D.~M., \& {Guiderdoni}, B. 1993, \mnras, 264, 201

\bibitem[{Kroupa(2001)}]{kroupa_2001}
Kroupa, P. 2001, \mnras, 322, 231

\bibitem[{{Larson}(1975)}]{larson_1975}
{Larson}, R.~B. 1975, \mnras, 173, 671

\bibitem[{Mannucci {et~al.}(2002)Mannucci, Pozzetti, Thompson, Oliva, Baffa,
  {et~al.}}]{mannucci_et_al_2002}
Mannucci, F., Pozzetti, L., Thompson, D., {et~al.} 2002, \mnras, 329, L57

\bibitem[{Moriondo {et~al.}(2000)Moriondo, Cimatti, \& Daddi}]{moriondo2000}
Moriondo, G., Cimatti, A., \& Daddi, E. 2000, \aap, 364, 26

\bibitem[{Moustakas {et~al.}(2003)Moustakas, Casertano, Conselice, Dickinson,
  Eisenhardt, {et~al.}}]{moustakas_et_al_2003}
Moustakas, L.~A., Casertano, S., Conselice, C., {et~al.} 2003, \apjl, submitted
  (astro-ph/0309187)

\bibitem[{Moustakas \& Somerville(2002)}]{moustakas_somerville_2002}
Moustakas, L.~A. \& Somerville, R.~S. 2002, \apj, 577, 1

\bibitem[{Pierini {et~al.}(2003)Pierini, Maraston, Bender, \&
  Witt}]{pierini_et_al_2003}
Pierini, D., Maraston, C., Bender, R., \& Witt, A.~N. 2003, \mnras, accepted
  (astro-ph/0309223)

\bibitem[{Pozzetti \& Mannucci(2000)}]{pm2000}
Pozzetti, L. \& Mannucci, F. 2000, \mnras, 317, L17 (PM00)

\bibitem[{Salpeter(1955)}]{salpeter1955}
Salpeter, E.~E. 1955, \apj, 121, 161

\bibitem[{Saracco {et~al.}(2003)Saracco, Longhetti, Severgnini, Della~Ceca,
  Mannucci, {et~al.}}]{saracco_et_al_2003}
Saracco, P., Longhetti, M., Severgnini, P., {et~al.} 2003, \aap, 398, 127

\bibitem[{Scalo(1998)}]{scalo1998}
Scalo, J. 1998, in ASP Conf. Ser. 142: The Stellar Initial Mass Function (38th
  Herstmonceux Conference), 201

\bibitem[{Scodeggio \& Silva(2000)}]{ss2000}
Scodeggio, M. \& Silva, D.~R. 2000, \aap, 359, 953

\bibitem[{Smail {et~al.}(2002)Smail, Owen, Morrison,
  {et~al.}}]{smail_et_al_2002}
Smail, I., Owen, F.~N., Morrison, G.~E., {et~al.} 2002, \apj, 581, 844 (S02)

\bibitem[{{Smith} {et~al.}(2002{\natexlab{a}}){Smith}, {Smail}, {Kneib},
  {Czoske}, {Ebeling}, {Edge}, {Pell{\' o}}, {Ivison}, {Packham}, \& {Le
  Borgne}}]{smith_et_al_2002a}
{Smith}, G.~P., {Smail}, I., {Kneib}, J.-P., {et~al.} 2002{\natexlab{a}},
  \mnras, 330, 1

\bibitem[{{Smith} {et~al.}(2002{\natexlab{b}}){Smith}, {Smail}, {Kneib},
  {Davis}, {Takamiya}, {Ebeling}, \& {Czoske}}]{smith_et_al_2002b}
{Smith}, G.~P., {Smail}, I., {Kneib}, J.-P., {et~al.} 2002{\natexlab{b}},
  \mnras, 333, L16

\bibitem[{{Smith} {et~al.}(2001){Smith}, {Treu}, {Ellis}, {Smail}, {Kneib}, \&
  {Frye}}]{smith_et_al_2001}
{Smith}, G.~P., {Treu}, T., {Ellis}, R., {et~al.} 2001, \apj, 562, 635

\bibitem[{{Soifer} {et~al.}(1999){Soifer}, {Matthews}, {Neugebauer}, {Armus},
  {Cohen}, {Persson}, \& {Smail}}]{soifer_et_al_1999}
{Soifer}, B.~T., {Matthews}, K., {Neugebauer}, G., {et~al.} 1999, \aj, 118,
  2065

\bibitem[{{Spinrad} {et~al.}(1997){Spinrad}, {Dey}, {Stern}, {Dunlop},
  {Peacock}, {Jimenez}, \& {Windhorst}}]{spinrad_et_al_1997}
{Spinrad}, H., {Dey}, A., {Stern}, D., {et~al.} 1997, \apj, 484, 581

\bibitem[{{Stiavelli} {et~al.}(1999){Stiavelli}, {Treu}, {Carollo}, {Rosati},
  {Viezzer}, {Casertano}, {Dickinson}, {Ferguson}, {Fruchter}, {Madau},
  {Martin}, \& {Teplitz}}]{stiavelli_et_al_1999}
{Stiavelli}, M., {Treu}, T., {Carollo}, C.~M., {et~al.} 1999, \aap, 343, L25

\bibitem[{Thompson {et~al.}(1999)Thompson, Beckwith, Fockenbrock,
  {et~al.}}]{thompson1999}
Thompson, D., Beckwith, S.~V.~W., Fockenbrock, R., {et~al.} 1999, \apj, 523,
  100

\bibitem[{{van Dokkum} \& {Stanford}(2003)}]{van_dokkum_et_al_2003}
{van Dokkum}, P.~G. \& {Stanford}, S.~A. 2003, \apj, 585, 78

\bibitem[{{White} \& {Frenk}(1991)}]{white_frenk_1991}
{White}, S.~D.~M. \& {Frenk}, C.~S. 1991, \apj, 379, 52

\bibitem[{Woosley \& Weaver(1995)}]{ww1995}
Woosley, S.~E. \& Weaver, T.~A. 1995, \apjs, 101, 181

\bibitem[{{Yan} \& {Thompson}(2003)}]{yan_thompson_2003}
{Yan}, L. \& {Thompson}, D. 2003, \apj, 586, 765

\bibitem[{{Zackrisson} {et~al.}(2001){Zackrisson}, {Bergvall}, {Olofsson}, \&
  {Siebert}}]{zackrisson_et_al_2001}
{Zackrisson}, E., {Bergvall}, N., {Olofsson}, K., \& {Siebert}, A. 2001, \aap,
  375, 814

\end{thebibliography}

\end{document}